# NONCOLLINEAR TOPOLOGICAL TEXTURES IN TWO-DIMENSIONAL VAN DER WAALS MATERIALS: FROM MAGNETIC TO POLAR SYSTEMS

Xiaoyan Yao,[*] Yu Wang, Shuai Dong[§]

*School of physics, Southeast University,*
*Nanjing, 211189, P. R. China*
[*] *yaoxiaoyan@seu.edu.cn*   [§] *sdong@seu.edu.cn*

In recent years, noncollinear topological textures have long gained increasing research attentions for their high values of both fundamental researches and potential applications. The recent discovery of intrinsic orders in magnetic and polar two-dimensional van der Waals materials provides a new ideal platform for the investigation of noncollinear topological textures. Here, we review the theoretical and experimental progresses on noncollinear topological textures in two-dimensional van der Waals materials in very recent years. During these years, magnetic skyrmions of both Bloch and Néel types have been observed experimentally in a few two-dimensional van der Waals materials and related heterostructures. Concurrently, more theoretic predictions basing on various mechanisms have been reported about different noncollinear topological textures in two-dimensional van der Waals materials, such as skyrmions, bimerons, anti-biskyrmions and skyrmionium, which are still waiting to be confirmed in experiments. Besides, noncollinear topological electric dipole orders have also been predicted in two-dimensional van der Waals materials. Taking advantage of the intrinsic two-dimensional nature and high integratability, the two-dimensional van der Waals materials will play an important role in the investigation on noncollinear topological textures in both magnetic and polar systems.

## 1. Introduction

Topology has long been an important part of physics research, and the noncollinear textures with topological untrivial property have ignited a growing interest in recent years, due to the novel physical phenomena and potential applications. Up to now, a variety of noncollinear magnetic textures have been proposed in theory and some of them have been observed in experiments, such as vortex, anti-vortex, skyrmion, anti-skyrmion, bimeron, biskyrmion and Hopfion.[1-5] In particular, magnetic skyrmions, have attracted a great deal of attention from both the academic and technological fields since the first experimental discovery in MnSi.[6] Their inherent stability ensured by the protected topology and the localized solitonic character make them technologically appealing for future information carriers in energy-efficient logic and memory devices. Skyrmionic magnetic texture is a natural excitation of ferromagnet in two spacial dimensions, carrying nontrivial topological charge (topological invariant serving as a fingerprint for a topological equivalence class) defined for two-dimensional (2D) structures.[7] It has been verified by experiments that the stability of skyrmions essentially depends on the dimensions of the system: the skyrmions are stable over a wide range of temperature and magnetic fields in the thin film, whereas they are relatively unstable and survive only in a narrow parameter region in the bulk.[8-10] Besides widely reported noncollinear magnetism, noncollinear electric dipole orders, such as flux-closure domains, polar vortices, and even room temperature polar-skyrmion have also been discovered in experiments.[11-15]

In the past decade, 2D van der Waals (vdW) materials have been significantly highlighted due to diverse novel phenomena over a broad range of electronic, magnetic, thermal, optical properties, and high integratability to produce appealing artificial heterostructures.[16] The 2D monolayers exfoliated from these 2D vdW materials are intrinsically superior at the nanoscale as compared with canonical three-dimensional (3D) materials. It is inspiring that long-range intrinsic magnetic orders have been discovered experimentally in the 2D vdW materials down to atomic layers recently, such as $Cr_2Ge_2Te_6$[17], $CrI_3$[18], $Fe_3GeTe_2$[19], $VSe_2$[20] and $MnSe_2$[21]. Even earlier, the robust ferroelectricity was also realized experimentally in the 2D vdW materials, such as $CuInP_2S_6$[22], $SnTe$[23] and $In_2Se_3$[24]. These 2D long-range orders have attracted considerable attention, and a lot of efforts have been put into the related investigations for their fundamental physics in low dimensions and their potential applications to multifunctional devices. These progresses provide an ideal platform to explore exotic noncollinear topological textures in 2D vdW materials.

In this article, we review very recent advances achieved in the investigations on noncollinear topological textures in 2D vdW materials, including both magnetic and polar systems. The experimental progresses are demonstrated first, and then the theoretical and computational predictions are discussed. Besides relatively more investigations on the noncollinear topological magnetism in 2D vdW materials, the reports about the noncollinear topological electric dipole order in 2D vdW structures remain rare, which will be discussed separately in the rear.

## 2. Noncollinear topological magnetic textures in 2D vdW materials

### 2.1. *Experimental progresses*

In 2019, topologically nontrivial magnetic textures were observed directly in experiments on exfoliated 2D vdW materials by Lorentz transmission electron microscopy (LTEM). The skyrmionic bubbles were first reported by Myung-Geun Han *et al*. in ferromagnetic insulator $Cr_2Ge_2Te_6$[25], and then by Bei Ding *et al*. in ferromagnetic metal $Fe_3GeTe_2$[26]. Both them possess centrosymmetric crystal structures and a strong magnetic anisotropy with an out-of-plane easy axis. The magnetic stripe domains as ground state transform into Bloch-type skyrmionic bubbles with an external magnetic field applied perpendicularly to the plane. These skyrmionic bubbles are stabilized by the competition between uniaxial magnetic anisotropy and magnetic dipole-dipole interactions. Moreover, a densely hexagonal lattice of skyrmion bubbles emerges by a field-cooling manipulation. Due to their topological stability, the skyrmion bubble lattices are stable to large field-cooling tilted angles, as shown in Fig. 1(a-d). And the size of bubbles is inversely scaled with external magnetic field as displayed in Fig. 1(e-h). Later, the zero-field-stabilized skyrmionic magnetic bubbles of Bloch-type were also observed in 20-nm-thick $Fe_3GeTe_2$ nanolayers at low temperature by LTEM and electron holography analyses.[27]

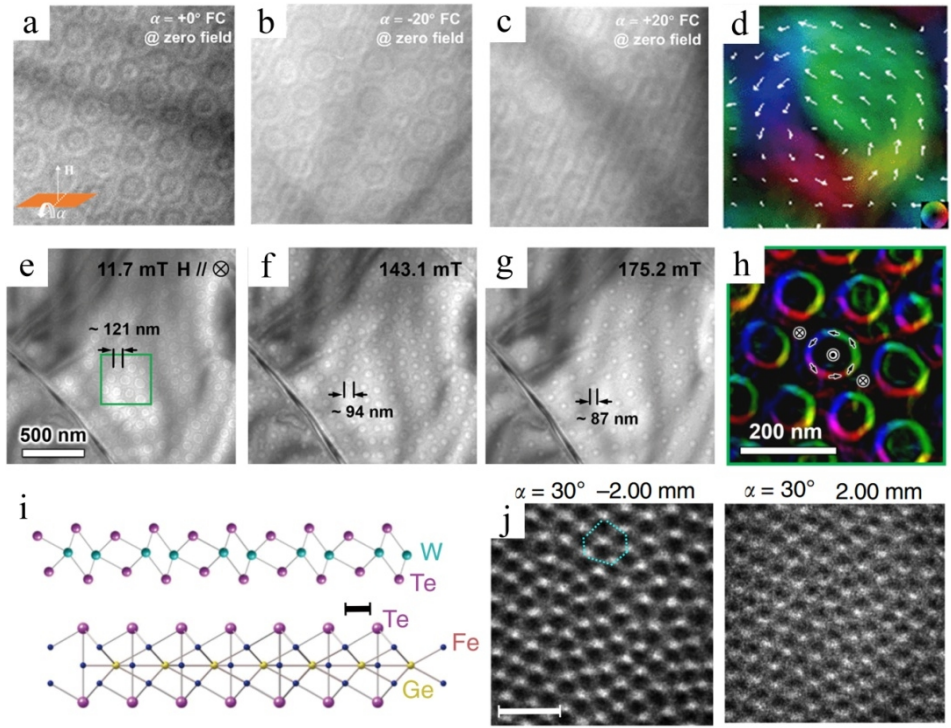

Fig. 1 (a-d) LTEM images of skyrmion bubbles in $Fe_3GeTe_2$ with a magnetic field (600 Oe) applied with rotation angles α, and an enlarged in-plane magnetization distribution map for a selected skyrmion bubble of Bloch-type is plotted in (d). Reprinted with permission from B. Ding *et al.*, *Nano Lett.* **20**, 868 (2020). Copyright 2019 American Chemical Society. (e-h) LTEM images showing the magnetic field dependence of skyrmionic bubble lattice in $Cr_2Ge_2Te_6$, and the detailed magnetization map demonstrating Bloch-type skyrmionic bubbles from the areas indicated by green box in (e) is presented in (h). Reprinted with permission from M.-G. Han *et al.*, *Nano Lett.* **19**, 7859 (2019). Copyright 2019 American Chemical Society. (i) Schematic graph for $WTe_2/Fe_3GeTe_2$ vdW heterostructure, on which the LTEM observation of skyrmion lattice is shown in (j). From under focus to over focus, the skyrmions transform from white above and dark below to the opposite contrasts, consistent with the Néel-type skyrmions. Scale bar: 500 nm. Reprinted with permission from Y. Wu *et al.*, *Nat. Commun.* **11**, 3860 (2020). Copyright 2020 The Author(s).

Nearly at the same time, magnetic skyrmions were also experimentally observed in the vdW heterostructures by LTEM and other high spatial resolution magnetic imaging techniques, and they tend to be Néel-type owing to the strong interfacial Dzyaloshinskii-Moriya interaction (DMI). Yingying Wu *et al.*[28] reported a lattice of Néel-type skyrmions at the $WTe_2/Fe_3GeTe_2$ interface, as show in Fig. 1(i) and (j). The large DMI energy from the interfacial coupling is ascribed to the broken inversion symmetry from Rashba spin-orbit coupling (SOC). Another experimental observation of Néel-type magnetic skyrmions was reported in $Fe_3GeTe_2$ on top of Co/Pd multilayers without external magnetic field.[29] In addition, Néel-type magnetic skyrmions and their lattice formation were also observed in $Fe_3GeTe_2$-based heterostructures[30], where the emergence of Néel-type skyrmions results from the significant DMI at the oxidized interfaces of $Fe_3GeTe_2$. Moreover, in a stacked $Fe_3GeTe_2$/h-BN vdW bilayer structure, individual skyrmion can be driven by using nanoseconds current-pulses.

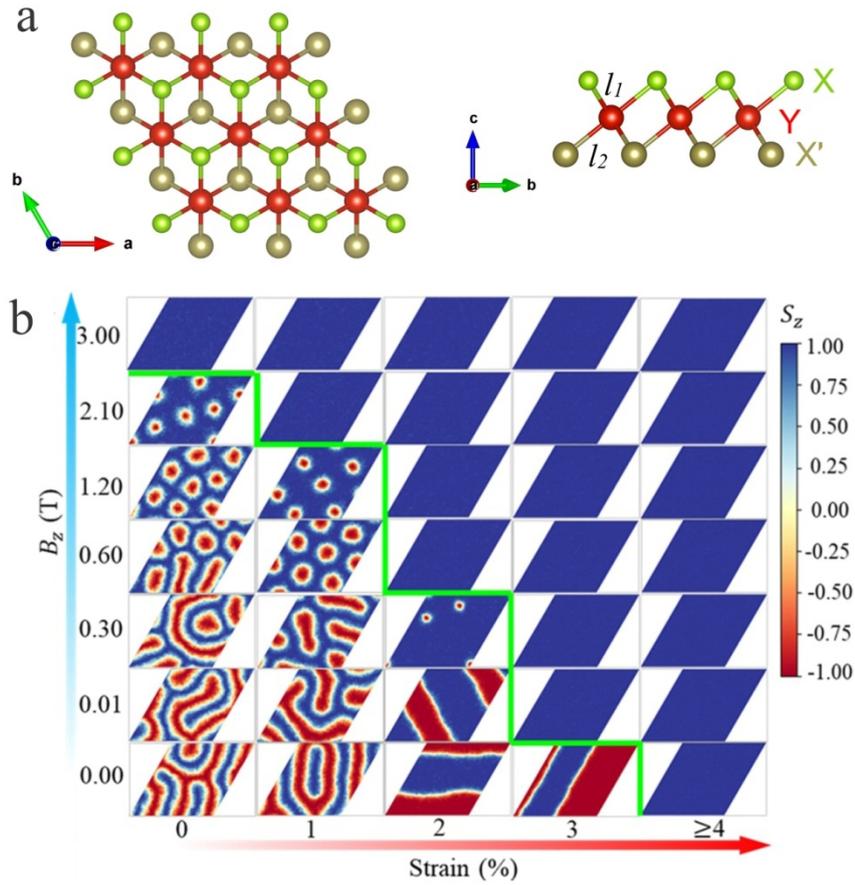

Fig. 2 (a) Top and side views of Janus monolayer of transition metal chalcogenides YXX' (Y=V, Cr, Mn; X, X'= S, Se, Te, X≠X'). (b) Spin textures for CrSeTe monolayer under increased tensile strains and external magnetic fields in 10 K. The color map indicates the out-of-plane spin component of Cr atoms. Reprinted with permission from Q. Cui *et al.*, *Phys. Rev. B* **102**, 094425 (2020). Copyright 2020 American Physical Society.

**2.2. *Theoretical predictions***

Besides the experiments mentioned above, many efforts have been made on theoretical and computational investigations to explore noncollinear topological magnetic textures in 2D vdW materials. A general strategy is to perform first-principles density functional theory (DFT) calculations on certain materials, and then atomistic spin modeling or micromagnetic simulation are carried out based on the parameters obtained from DFT calculations to present magnetic configurations. The detailed investigations are classified below according to different mechanisms for noncollinear topological magnetic textures.

2.2.1. *DMI induced by Janus structure*

Up to now, magnetic skyrmions, mostly observed in experiments of noncentrosymmetric materials or interfacial symmetry-breaking heterostructures, are primarily induced by the DMI. For the presence of DMI, in addition to the strong SOC, the system is required to have broken inversion symmetry. However, most of the 2D vdW magnets possess centrosymmetric structure, so the DMI is absent. Following this line of thought, different methods have been proposed to break the inversion symmetry artificially to generate DMI. One desirable situation is 2D magnet with inherent inversion asymmetry and intrinsic DMI, which can be effectively realized by fabricating Janus monolayers. Janus monolayer, with different atoms occupying top and bottom layers respectively, has out-of-plane structural asymmetry, and thus is predicted to possess considerable DMI and promising to create magnetic skyrmions.

The experiments have demonstrated that Janus monolayers of transition metal chalcogenides, e.g., nonmagnetic MoSSe, can be synthesized.[31-33] Theoretically, several 2D magnetic Janus monolayers of transition metal chalcogenides YXX' (Y=V, Cr, Mn; X, X'=S, Se, Te, X≠X') are investigated systematically, where large intrinsic DMI and stable skyrmions were predicted.[34-37] As illustrated in Fig. 2(a), Y atoms form a hexagonal network sandwiched by two atomic planes of different chalcogen atoms X and X'. The difference between X and X' on the opposite sides of Y with different bond lengths $l_1$ and $l_2$ breaks the inversion symmetry, thus allowing the DMI between the Y ions. Jiaren Yuan *et al.*[34] and Jinghua Liang *et al.*[35] predicted significant intrinsic DMI in MnSTe and MnSeTe monolayers with a strong SOC originated from the Te atoms. Both them behave the perpendicular magnetic anisotropy with an out-plane easy axis. At low temperatures the ground states can transform from ferromagnetic states with wormlike magnetic domains into the skyrmion states by applying an external magnetic field. Moreover, the sub-50-nm Néel skyrmions could be stabilized in both monolayers at zero magnetic field, and the size could be shrunk to sub-10 nm by an external magnetic field. Qirui Cui *et al.*[36] reported that CrSeTe monolayer with large DMI also hosts Néel domain wall, which turn into skyrmion states by applying an external magnetic field, and the diameter and density of skyrmions can be tuned by strain, as shown in Fig. 2(b). Slimane Laref *et al.*[37] predicted that VSSe possesses a weak out-of-plane uniaxial anisotropy, whereas VSeTe exhibits a strong easy-plane anisotropy. DMI is large enough to stabilize non-trivial chiral textures for both them. In VSeTe monolayer, an asymmetrical bimeron lattice state is stabilized upon in-plane magnetic field, whereas skyrmion lattice is formed under out-of-plane magnetic field. Skyrmion lattice is also stabilized in VSSe monolayer with out-of-plane field applied.

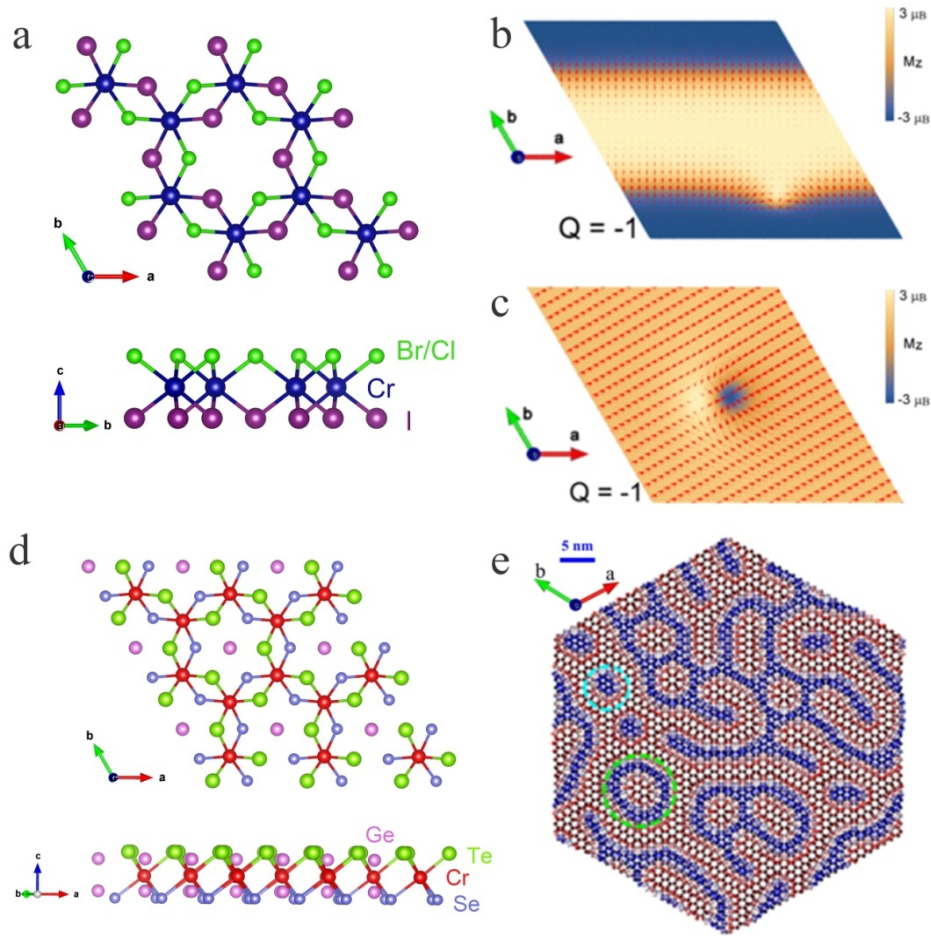

Fig. 3 (a) The side and top views of Cr(I,X)$_3$ (X =Br, Cl). (b) One domain wall skyrmion at domain boundary obtained for Cr(I,Br)$_3$. (c) One bimeron upon an out-of-plane magnetic field of 0.8 T obtained for Cr(I,Cl)$_3$. The color map applies to the out-of-plane component of the spin vectors, while the arrows characterize the in-plane components of the spin vectors. Reprinted with permission from C. Xu *et al.*, *Phys. Rev. B* **101**, 060404 (2020). Copyright 2020 American Physical Society. (d) The top and side views of CrGe(Se,Te)$_3$ monolayer. (e)The spin texture of the CrGe(Se, Te)$_3$ monolayer at 0.1K. The skyrmion and skyrmionium is highlighted with cyan and green circle, respectively. The red and blue colors denote magnetic moments pointing up and down. Reprinted with permission from Y. Zhang *et al.*, *Phys. Rev. B* **102**, 241107 (2020). Copyright 2020 American Physical Society.

The strategy of Janus monolayer was also been applied to other 2D ferromagnetic vdW materials. Changsong Xu *et al.*[38] proposed the fabrication of Janus monolayers Cr(I, X)$_3$ (X =Br, Cl), where the Cr atoms form a honeycomb network, which are covalently bonded to the underlying I ions and top-layer Br or Cl atoms, as illustrated in Fig. 3(a). The asymmetry between the top and bottom layers allows the strong DMI between the Cr ions. It is found that Cr(I, Br)$_3$ with a mild out-of-plane anisotropy can intrinsically host metastable domain wall skyrmion phase, while a bimeron state can be stabilized in Cr(I, Cl)$_3$ with a weak in-plane anisotropy by applying an out-of-plane magnetic field, as depicted in Fig. 3(b) and (c). Yun Zhang *et al.*[39] predicted another stable Janus monolayer CrGe(Se,Te)$_3$, as shown in Fig. 3(d), exhibiting a strong DMI and strong out-of-plane anisotropy. Consequently, nanometric Néel-type skyrmions and skyrmioniums (with topological charge $Q$=0) can spontaneously form in the absence of magnetic field, as demonstrated in Fig. 3(e). It was unveiled that the subtle competition between DMI and a moderate exchange frustration results in the stabilization of the skyrmionium.

### 2.2.2. *DMI induced by polar displacements*

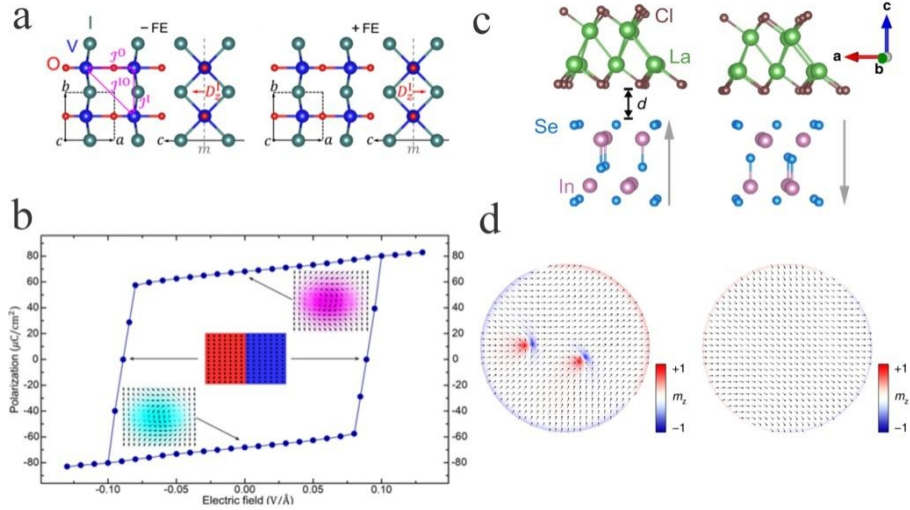

Fig. 4 (a) The ferroelectric phases with opposite polarizations (±FE) for multiferroic VOI$_2$ monolayer, where $D_z^I$ presents the direction of DMI. And (b) the polarization-electric field loop from DFT with the insets indicating the spin patterns at different states with opposite topological charge. Reprinted with permission from C. Xu *et al.*, *Phys. Rev. Lett.* **125**, 037203 (2020). Copyright 2020 American Physical Society. (c) The side views of LaCl/In$_2$Se$_3$ heterostructure, where the gray arrows represent the polarization direction of In$_2$Se$_3$, namely P+ and P−. (d) The top views of the micromagnetic simulation for LaCl/P+ (left) and LaCl/P− (right). Reprinted with permission from W. Sun *et al.*, *Nat. Commun.* **11**, 5930 (2020). Copyright 2020 The Author(s).

Besides Janus monolayers synthesized artificially, inversion asymmetry may also be found in polar structures, where the electric polarization is expected to play an important role in tuning DMI and then to control magnetic texture. It is predicted by Jinghua Liang *et al.*[40] that the structure inversion asymmetry induced by vertical ferroelectric displacement in 2D multiferroics could lead to a significant DMI. Sub-10-nm skyrmions stabilized by the DMI with perpendicular magnetic anisotropy could be realized in CrN monolayer. Moreover, the effective electric control of the strength and chirality of the DMI and thereby the magnetic textures can be obtained. A little early, magnetic bimerons were predicted by Changsong Xu *et al.* to exist in multiferroic $VOI_2$ monolayer with both ferromagnetism and ferroelectricity.[41] In the ferroelectric phase, inversion symmetry is broken by the polar displacements of V, which then results in a significant DMI with opposite directions according to opposite polarizations, as illustrated in Fig. 4(a). The bimerons with different topological charges are stabilized by the opposite DMIs and in-plane anisotropy. Furthermore, a novel mechanism of switching topological charge by electric-field in a controllable and reversible fashion was proposed, through the mediation of coupled electric polarization and DMI, as shown in Fig. 4(b). Magnetic bimerons were also predicted in the atomic layer-thick vdW $LaCl/In_2Se_3$ multiferroic heterostructure.[42] Here 2D ferromagnetic LaCl monolayer is metallic with in-plane magnetization, while $In_2Se_3$ is 2D ferroelectric layer with controllable out-of-plane spontaneous polarization, as illustrated in Fig. 4(c). Together with a strong SOC from La-5d orbital, the ferroelectric polarization originating from the $In_2Se_3$ layer breaks the spatial inversion symmetry in the adjacent LaCl, which gives rise to a strong DMI in this heterostructure. Different from the situation in $VOI_2$ mentioned above, asymmetric magnetic behavior appears, that is, bimerons emerge in a ferroelectric state, but disappear when the polarization is reversed, as shown in Fig. 4(d). Therefore these bimerons can be generated or annihilated by switching ferroelectric polarization in $In_2Se_3$ by an external electric field.

Besides intrinsic ferroelectric displacements in 2D multiferroics, DMI can also result from the ionic displacements induced by external electrical field. Jie Liu *et al.*[43] and Aroop K. Behera *et al.*[44] proposed an application of out-of-plane electrical field on $CrI_3$ ferromagnetic monolayer to produce sub-10-nm Néel-type magnetic skyrmions. While an electric field applied in the vertical direction to the plane of $CrI_3$, the opposite movements of I planes and Cr plane are caused along the direction of the field. The lattice distortion with unequal-distanced I planes from Cr plane leads to the breaking of inversion symmetry. The induced Rashba-type SOC gives rise to a net DMI in the system. The stable Néel-type skyrmions arise due to a competition between magnetocrystalline anisotropy, DMI, and the magnetic field. However, the electric field required to induce a sufficient DMI is in an extremely large order of V/nm.

### 2.2.3. *Magnetic frustration or dipole-dipole interaction in centrosymmetric materials*

While the DMI is widely studied as a main road to noncollinear magnetic textures in the noncentrosymmetric materials, some other mechanisms are discussed for centrosymmetric systems, even also for 2D vdW materials. Very recently, Danila Amoroso *et al.*[45] predicted a novel noncollinear topological texture and a field-induced topological transition in the vdW $NiI_2$ monolayer. $NiI_2$ is a magnetic semiconductor with centrosymmetric structure, where magnetic cations form a frustrated triangular net, as plotted in Fig. 5(a). The competing ferromagnetic and antiferromagnetic interactions result in strong magnetic frustration. As presented in Fig. 5(c), a triangular lattice of anti-biskyrmions characterized by a high topological charge ($|Q|=2$) can be spontaneously stabilized as ground-state with unique topology and chirality. As a perpendicular magnetic field is applied, a topological transition to a conventional Bloch-type skyrmion ($|Q|=1$) lattice (Fig. 5(d)) appears, as also shown in Fig. 5(b). It was revealed that the anisotropic part of the short-range symmetric exchange interaction, combined with the exchange frustration, in absence of DMI and Zeeman interactions, leads to the predicted topological magnetic texture. In addition, the existence of meron-like spin textures was predicted in vdW monolayer $CrCl_3$ with weak SOC and easy-plane anisotropy.[46] The magnetic dipole-dipole interactions play an important role in generating these merons with pairing character, which are robust against external fields.

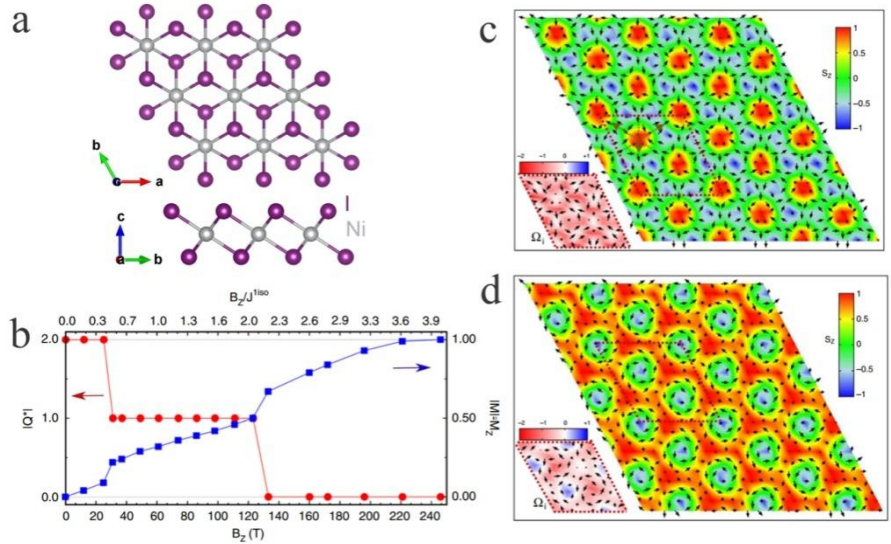

Fig. 5 (a) The side and top views of $NiI_2$ monolayer. (b) Evolution of $|Q|$ and magnetization as a function of the magnetic field $B_z$ at T = 1K. Accordingly, the spin configurations of the anti-biskyrmion ($|Q|=2$) and Bloch-skyrmion ($|Q|=1$) lattices are displayed in (c) and (d) respectively. Reprinted with permission from D. Amoroso *et al.*, *Nat. Commun.* **11**, 5784 (2020). Copyright 2020 The Author(s).

2.2.4. *Moiré pattern*

Aiming at 2D heterostructures, a new mechanism of noncollinear magnetic texture was proposed by Qingjun Tong *et al.*[47], that is, skyrmions can be induced by the moiré pattern in 2D vdW heterostructures, which could be constructed by a ferromagnetic monolayer ($CrX_3$, X = Cl, Br, and I) on an antiferromagnetic substrate ($MnPX_3$, X = S, Se, Te), or $CrI_3$ homolayers. The interlayer magnetic coupling modulated by the different local atomic sites induces skyrmions, as illustrated in Fig. 6. The moiré skyrmions with opposite topological charge are degenerate, and they are trapped at an ordered array with the moiré periodicity. With relatively strong interlayer coupling or large moiré periodicity, skyrmion lattices are the ground state, which can be dramatically tuned by small interlayer translation and strain. With relatively weak interlayer coupling or small moiré period, metastable skyrmion excitations exist on the ferromagnetic ground state, which can be generated and erased by reversing local magnetization, and can be moved between the ordered trapping sites by current pulses.

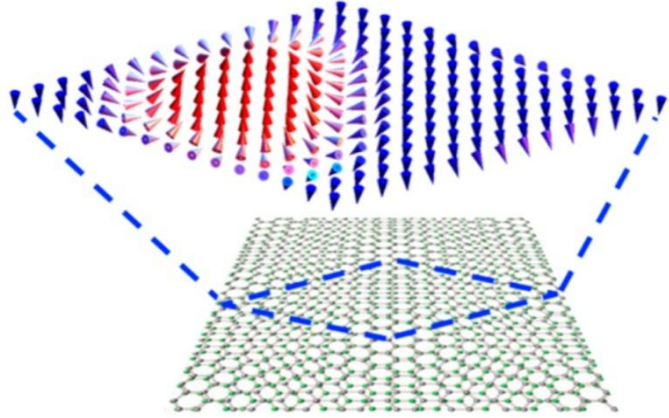

Fig. 6 Skyrmion formation from the moiré pattern in 2D vdW heterostructures. Reprinted with permission from Q. Tong *et al.*, *Nano Lett.* **18**, 7194 (2018). Copyright 2018 American Chemical Society.

## 3. Noncollinear topological electric dipole textures in 2D vdW materials

Noncollinear topological textures of electric dipoles have been reported in some investigations of both experiment and theory, but it is rarely discussed in 2D vdW materials. In 2019, Ling-Fang Lin *et al.*[48] predicted by DFT calculations that the monolayer $MO_2X_2$, *i.e.* $WO_2Cl_2$ and $MoO_2Br_2$, are promising 2D polar materials to display intrinsic noncollinear ferrielectricity. As plotted in Fig. 7(a) and (b), the spatially loose "square" lattice of $MO_2X_2$ is advantageous for polar distortions. The $d_0$ rule, *i.e.*, the formation of coordination bonds between M's empty d orbitals and O's 2p orbitals, is the driving force for the polar distortions. Then the frustration between ferroelectric and antiferroelectric soft modes generates intrinsically noncollinear electric dipole moments, as illustrated in Fig. 7(c). There are four domains: A+, A−, B+, and B− corresponding to four degenerate lowest-energy wells as shown in Fig. 7(d), and the favored domain walls are between A+/A−, A+/B+, B+/B−, and A−/B−. If the boundaries between A+ and A- (B+ and B-) domains are marked by ferroelectric (FE) domain walls, and those between A+ and B+ (A- and B-) domains by antiferroelectric (AFE) domain walls, the atomic-scale dipole vortices and antivortices form at the FE domain walls, and topological $Z_2 \times Z_2$ antiphase domain vortices (antivortices) emerge at the cross points of FE and AFE domain walls, as displayed in Fig. 7(e).

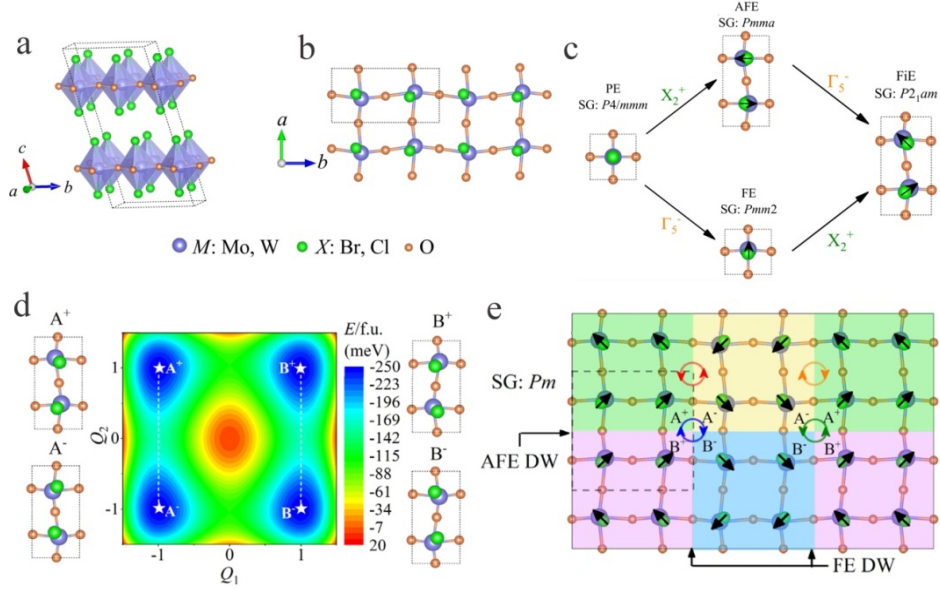

Fig. 7 (a) Structure of $MO_2X_2$. (b) Top view of a monolayer (dashed lines indicate a unit cell). (c) Structural transitions of a $MO_2X_2$ monolayer, i.e. from paraelectric (PE) phase to stable ferrielectric (FiE) ground state via intermediate ferroelectric (FE) or antiferroelectric (AFE) phases. $X_2^+$ and $\Gamma_5^-$ represent two unstable distortion modes. (d) Contour plot of the energy with varying $Q_1$ and $Q_2$ (amplitudes of $X_2^+$ and $\Gamma_5^-$ modes). (e) Schematic of four-colored domains. Red (orange) circles denote a vortex (antivortex) at the FE domain walls. The blue (green) circle marks a topological $Z_2 \times Z_2$ antiphase domain vortex (antivortex). Reprinted with permission from L.-F. Lin *et al.*, *Phys. Rev. Lett.* **123**, 067601 (2019). Copyright 2019 American Physical Society.

## 4. Summary

In summary, we review very recent progresses in the field of noncollinear topological textures in 2D vdW materials, including experimental and theoretical investigations in both magnetic and polar systems. During the short three years since 2018, magnetic skyrmions of both Bloch and Néel types have been observed experimentally in a few 2D vdW materials and related heterostructures. Concurrently, more theoretical predictions about different noncollinear topological magnetic textures have been proposed in 2D vdW materials, basing on various mechanisms, such as DMI, frustration and moiré pattern, which are still waiting to be confirmed in experiments. On the other hand, noncollinear topological texture of electric dipoles has been predicted in ferrielectric 2D vdW material, although the reports on this aspect for polar systems are relatively rare.

Only during the past five years, the intrinsic ferromagnetism and ferroelectricity were discovered in 2D vdW materials. Therefore, this is a very young topic with great interest from the viewpoints of both fundamental research and technological applications. 2D vdW materials with extremely small thickness and unique 2D physical nature provide a perfect platform for exploring noncollinear topological textures. Further taking advantage of their high controllability and integratability, 2D vdW heterostructures can be constructed, offering a powerful framework for nanoscale-device engineering. The widely tunable properties of 2D vdW materials and heterostructures will enable the formation of diverse topological textures, and further enable tunable manipulation on these topological textures via applying various external stimuli, varying the thickness, the interfacial symmetry, the magnetocrysalline anisotropy, and so on.

## Acknowledgments

This work is supported by the research grants from the National Natural Science Foundation of China (Grant No. 11834002).